# Operational and Exploration Requirements and Research Capabilities for SEP Environment Monitoring and Forecasting

Viacheslav M Sadykov[1] and Petrus Martens[2] and Dustin Kempton[3] and Rafal Angryk[4] and Berkay Aydin[5] and Jessica Hamilton[6] and Griffin Goodwin[7] and Aatiya Ali[8] and Sanjib K C[9] and Rimsha Syeda[10]
*Georgia State University (GSU), Atlanta, GA 30303*

Irina Kitiashvili[11]
*NASA Ames Research Center (NASA ARC), Moffett Field, CA 94035*

Kathryn Whitman[12]
*NASA Johnson Space Center (NASA JSC), Houston, TX 77058*

Alexander Kosovichev[13]
*New Jersey Institute of Technology (NJIT), Newark, NJ 07102*
*NASA Ames Research Center (NASA ARC), Moffett Field, CA 94035*

Kimberly Moreland[14]
*Cooperative Institute for Research in Environmental Sciences (CIRES), Boulder, CO 80309*

Manolis Georgoulis[15]
*Johns Hopkins University Applied Physics Laboratory (JHU APL), Laurel, MD 20723*

Ming Zhang[16]
*Florida Institute of Technology (FIT), Melbourne, FL 32901*

Azim Ahmadzadeh[17]
*The University of Missouri–St. Louis (UMSL), St. Louis, MO 63121*

Ronald Turner[18]
*Analytic Services Inc. (ANSER), Arlington, VA 22203*

---

[1] Assistant Professor, Physics & Astronomy Dept, GSU, 25 Park Place NE, Atlanta, GA 30303
[2] Professor, Physics & Astronomy Dept, GSU, 25 Park Place NE, Atlanta, GA 30303
[3] Research Professor, Dept of Computer Sciences, GSU, 25 Park Place NE, Atlanta, GA 30303
[4] Distinguished Professor, Dept of Computer Sciences, GSU, 25 Park Place NE, Atlanta, GA 30303
[5] Assistant Professor, Dept of Computer Sciences, GSU, 25 Park Place NE, Atlanta, GA 30303
[6] PhD Student, Physics & Astronomy Dept, GSU, 25 Park Place NE, Atlanta, GA 30303
[7] PhD Student, Physics & Astronomy Dept, GSU, 25 Park Place NE, Atlanta, GA 30303
[8] PhD Student, Physics & Astronomy Dept, GSU, 25 Park Place NE, Atlanta, GA 30303
[9] PhD Student, Physics & Astronomy Dept, GSU, 25 Park Place NE, Atlanta, GA 30303
[10] PhD Student, Dept of Computer Sciences, GSU, 25 Park Place NE, Atlanta, GA 30303
[11] Research AST, NASA Advanced Supercomputing Division, NASA ARC, Moffett Field, CA 94035
[12] Senior Research Scientist, Space Radiation Analysis Group, NASA JSC, 2101 NASA Parkway, Houston, TX 77058
[13] Professor, Physics Department, NJIT, 161 Warren St., Newark, NJ 07102
[14] Associate Scientist, CIRES, 1665 Central Campus Mall, 216 UCB, Boulder, CO 80309
[15] Senior Professional Staff, JHU APL, 11100 Johns Hopkins Road, Laurel, MD 20723
[16] Professor, Dept of Aerospace, Physics and Space Sciences, FIT, 150 W University Blvd, Melbourne, FL 32901
[17] Assistant Professor, Dept of Computer Science, UMSL, 311 Express Scripts Hall, St. Louis, MO 63121
[18] Distinguished Analyst, ANSER, 4040 Wilson Boulevard, Arlington, VA 22203

**Mitigating risks posed by solar energetic particles (SEPs) to operations and exploration in space and Earth's atmosphere motivates the development of advanced, synergistic approaches for monitoring, modeling, and analyzing space weather conditions. The consequences of SEPs and their interactions with the near-Earth space environment are numerous, including elevated radiation levels at aviation altitudes during major events, satellite damage, and health risks to astronauts, resulting in economic impacts and potential hazards for space exploration. This contribution will present a high-level overview of the operational requirements and research capabilities for SEP event environment monitoring and forecasting that were highlighted during a workshop at Georgia State University, held on October 16-19, 2024. Specifically, it summarizes the presented activities concerning the following: (1) Identifying needs for SEP event forecasting and nowcasting, including practical forecast timeframes; (2) Reviewing availability and coverage of the current observational data and identifying tangible data resources for research, operations and the R2O2R loop; (3) Mapping existing forecast capabilities and identifying meaningful modeling advances for research and operations.**

## Acronyms and Nomenclature

| | | |
|---|---|---|
| SEP | = | Solar Energetic Particle |
| CME | = | Coronal Mass Ejection |
| NSF | = | National Science Foundation |
| NASA | = | National Aeronautics and Space Administration |
| NOAA | = | National Oceanic and Atmospheric Administration |
| SRAG | = | Space Radiation Analysis Group |
| AR | = | Active Region |
| SC | = | Solar Cycle |
| EVA | = | Extravehicular Activity |
| ML | = | Machine Learning |
| DL | = | Deep Learning |
| MeV | = | Mega electron Volt |
| GeV | = | Giga electron Volt |
| pfu | = | particle flux unit |
| SOC | = | Scientific Organizing Committee |
| LOC | = | Local Organizing Committee |
| CCMC | = | Community Coordinated Modeling Center |
| SPHINX | = | Solar Particles in the Heliosphere validation INfrastructure for SpWx |

## I. Introduction to SEP Monitoring and Forecasting

Solar energetic particles (SEPs) are defined as the sudden enhancements of the fluxes of energetic protons, electrons, and heavier ions over the slowly varying galactic cosmic ray background observed with in-situ measurements in the Heliosphere and Geospace. There are various consequences of SEPs and their interaction with the near-Earth environment, including an increase in the radiation levels at aviation altitudes during the strongest events (Kataoka et al., 2018) corresponding economic impacts (e.g., changes in flight path length, time, and fuel consumption, Saito et al., 2021), damage to satellites, impacts on astronauts' health (Martens, 2018), and potential dangers for space exploration. Therefore, SEP events represent one of the most important components and agents of "space weather" — a term that encompasses the interaction of global solar activity and its manifestations in transient events with terrestrial and, more recently, with planetary environments. The importance of forecasting space weather was recognized at the United States governmental level by Promoting Research and Observations of Space Weather to Improve the Forecasting of Tomorrow Act (or PROSWIFT Act[3]) becoming a public law in October 2020, along with the

---

[3] https://www.congress.gov/bill/116th-congress/senate-bill/881

Implementation Plan of the National Space Weather Strategy and Action Plan[4] in December 2023 that emphasizes the need for coordination among the federal government, academia, the private sector, and international partners.

Among the particles constituting SEPs, the high-energy protons (with energies ranging from a few MeVs up to ~10 GeV) are often of primary importance for monitoring and prediction, given their multi-fold impact on the Geospace environment and space exploration. Over the last decade, there have been significant advances in SEP-related research, including efforts to develop comprehensive SEP catalogs (Robinson et al., 2018, Rotti et al., 2022), computationally-intensive physics-based approaches for modeling of SEP properties (Li et al., 2021, Zhao et al., 2020), and the involvement of novel machine learning (ML) techniques for SEP prediction (Ali et al., 2024, Kasapis et al., 2022, Rotti et al., 2023, 2024, O'Keefe et al., 2024). Current achievements in different scientific and technological areas make it possible to integrate interdisciplinary expertise and search for ways to enhance the understanding of SEP initiation and propagation. This can enable targeted, near-real-time monitoring of the radiation environment and operational forecasts of radiation hazards.

Nevertheless, mitigation of risks from SEPs in operations and exploration in space and Earth's atmosphere remains a challenging problem. There are multiple reasons for this, and we will illustrate some of them here. For example, for some SEP events, it is impossible to identify specific associated drivers (e.g., flare or coronal mass ejection, CME), and sometimes there is no particular associated Active Region (AR) on the visible side of the solar surface where the SEP event could have originated. For example, among the 433 SEP events of Solar Cycles 22-24 listed in the catalog of Rotti et al. (2022), 50 do not have the location of the host AR or flare identified. Another point is that SEP events are rare compared to CME or solar flare events, leading to a strong class imbalance in the data for the binary prediction problem. For example, the number of days with enhanced proton flux (with proton energies of $\geq 10$ MeV and related particle fluxes $\geq 10$ particle flux units, pfus) to the number of days with no enhanced flux is ~1/23 for Solar Cycles 22-24 (Ali et al., 2024). The class-imbalance issues are escalated even further if the SEP events with higher energy or flux thresholds are considered. The spatiotemporal variations of SEPs introduce an additional dimension to their analysis. SEPs can be observed from multiple vantage points, with their properties depending on the observational location (Richardson et al., 2018). The fluxes of the same SEP event detected at different heliospheric locations can even depend on particular CME characteristics, such as speed or width. With advances in space exploration, forecasting and analyzing radiation environments on the Moon and Mars have become increasingly important (Zhang et al., 2023, 2024). Ultimately, SEPs and their significance for space weather extend beyond the Sun-Earth line, presenting a broader heliospheric challenge.

Therefore, the problem of SEP monitoring and forecasting remains and requires a coordinated effort. To achieve progress, one needs to understand the current predictive capabilities of the research community and their alignment with operational/exploration needs, as well as to encourage and enhance dialogue between the operations/exploration and research communities on the SEP prediction problem. To facilitate progress, the authors of this contribution proposed and successfully conducted a workshop that brought together representatives of the operational and exploration agencies, data management and analysis experts, and researchers with experience developing SEP forecasting models.

This contribution is structured as follows. Section II describes the "Operational and Exploration Requirements and Research Capabilities for SEP Environment Monitoring and Forecasting" workshop that was conducted on October 16-19, 2024, at Georgia State University. Sections III-V overview the primary presentation topics and discussion points. Section VI concludes the contribution. The discussion topics presented in this contribution reflect the collective input of the workshop attendees, whose active participation and expertise significantly contributed to the content.

## II. Structure of the Workshop

**A. Workshop Goal and Objectives.**

The workshop "Operational and Exploration Requirements and Research Capabilities for SEP Environment Monitoring and Forecasting" was held on October 16-19, 2024, at Georgia State University (Atlanta, Georgia, USA) and targeted challenges associated with predicting of SEP events. The primary goal of the workshop was to highlight the research advances in SEP modeling and forecasting, to understand the related operational and exploration requirements, and to determine the supporting observational data availability and preparation needs. To achieve this goal, we enriched the dialogue between the research community and the exploration and operation entities by 1) conveying a clear message from operations and exploration entities to the research community on addressing specific needs and forecasting certain quantities of interest, 2) understanding the scope of current observations, ongoing data reprocessing and organization efforts, and identifying existing observational gaps, and 3) overviewing the prediction/modeling capabilities and limitations that the community currently faces, defining the research priorities to support forecasting interests.

---

[4]https://www.whitehouse.gov/wp-content/uploads/2023/12/Implementation-Plan-for-National-Space-Weather-Strategy-12212023.pdf

The specific objectives of the workshop were the following:
- Identify the needs and requirements of operational and exploration agencies and other users in terms of quantities related to the energetic particle content to be forecast or nowcast, and the corresponding timeframes of the forecasts;
- Revise the current observational data availability and coverage, the related data preparation and categorization efforts, and identify the data-related resources in support of the requirements above from the research and operations perspectives;
- Highlight the current modeling and forecasting capabilities of the community and identify models that are potentially tunable for predicting quantities of interest, as well as emphasize research directions allowing to address such predictions in the future.

**B. Workshop Logistics.**

The workshop was conducted in a hybrid format (providing options for in-person and remote participation). It was organized and coordinated by the Scientific Organizing Committee (SOC) and Local Organizing Committee (LOC), which is a standard way to organize the logistics of scientific conferences. SOC has provided scientific coordination of the workshop (workshop advertisement and announcements, program/agenda preparation, identification of the potential invited speakers, revision of the abstracts, and chairing duties). LOC has provided the logistic coordination of the workshop (communication with the participants, including transportation and accommodation, and preparation of the venue). SOC and LOC members are the co-authors of this contribution.

The duration of the hybrid workshop was 3.5 days, with about 2/3 of each workshop day dedicated to presentations and about 1/3 of each day to the open discussion sessions. The last half-day of the workshop was fully dedicated to discussions. Each day started with a presentation by the keynote speaker (40-minute talk including questions), followed by 12-14 contributed talks (20-minute talk including questions).

**C. Demographics of Participants.**

The workshop had 109 attendees. Among the attendees, 46 participants attended at least one day of the workshop in-person, and 63 joined online (Figure 1). This exceeded the target number of attendees assumed during the workshop planning phase, which was between 50 and 80, with an equal split between online and in-person attendees. The final agenda of the workshop is available at https://sites.google.com/view/sepmeeting2024/meeting-agenda.

Figure 2 illustrates the demographics of the workshop attendees depending on their career stages. The 'Other' category included attendees who self-identified differently from the default categories, such as representatives of funding agencies or governmental employees. As one can see, more than one-third of the participants (37) identified themselves as graduate or undergraduate students, and 8 were postdoctoral researchers. Thus, the workshop provided a great opportunity for early career colleagues to share and discuss their achievements.

### III. Identification of Needs and Requirements for SEP Monitoring and Forecasting.

To facilitate a transparent transition from research-driven developments to operational forecasting, it was essential to conduct a comprehensive overview of the current needs and capabilities in Solar Energetic Particle (SEP) monitoring and forecasting. There were presentations by space weather forecast stakeholders (operations, exploration, and advisory agencies) targeted at describing the operational or advisory activities and overviewing the current and expected requirements for forecasting the SEP environment and the efforts taken to address these requirements (e.g., Bain et al., 2021). This included the overview of operations and requirements from the NOAA Space Weather Prediction Center and NASA's Moon to Mars Space Weather Analysis Office. The workshop covered a wide range of aspects related to the space weather monitoring and SEP real-time forecasting activities by NASA Space Radiation Analysis Group (MagPy SEP forecasting model), the NASA Community Coordinated Modeling Center (CCMC; SEP scoreboard), near real-time CME alerts (ground-based coronagraph K-Cor), and the relativistic electron-based forecasting system for solar energetic protons (RELEASE), as well as the overviews of the Space Weather activities from NSF and NASA.

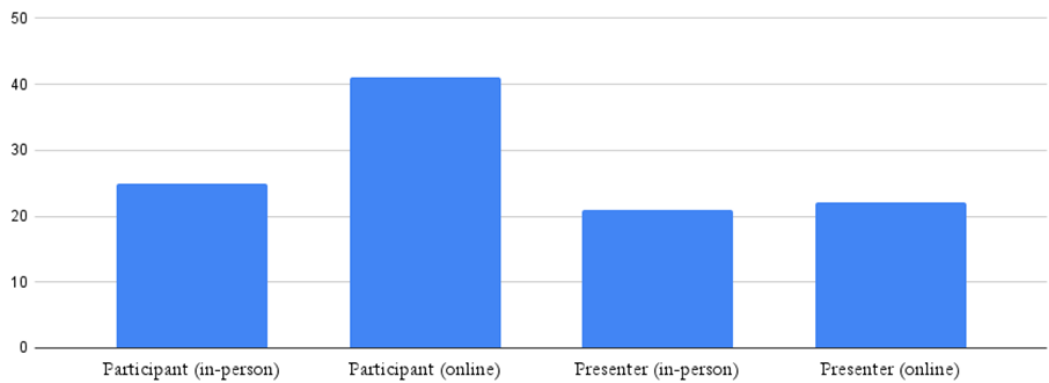

**Figure 1.** *Illustration of the number of workshop attendees with respect to their participation modes (presenter or participant, online or in-person). "Participant" in this figure refers to the attendee who did not present but participated in other workshop activities (such as discussions).*

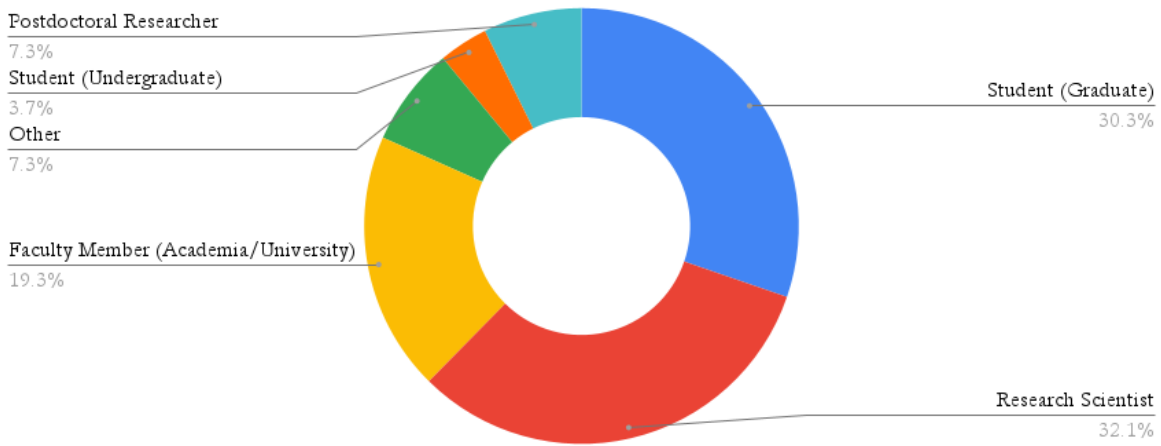

**Figure 2.** *Demographics of the "Operational and Exploration Requirements and Research Capabilities for SEP Environment Monitoring and Forecasting" workshop in terms of participant career status.*

It is crucial to have a clear understanding of the forecast-specific prediction data products and temporal requirements (e.g., forecast timescales and the energies of SEP protons) to enable an efficient transition from data processing and model development to reliable operational forecasting. In particular, it was emphasized that for the purposes of mission planning, launch commitment, and extravehicular activity (EVA), the 24-hour and 6-hour forecast time windows for the All-Clear forecast of > 100 MeV > 1 pfu protons, along with > 10 MeV > 10-100 pfu thresholds, are of interest. For spaceflights, when astronauts are required to seek shelter from the ≥ 300 MeV protons, the warning time must be at least ~10-30 minutes. The forecast requirements and event scales of interest can also be found for individual entities, such as the Space Radiation Analysis Group[5] (SRAG) or NOAA (S-scales[6]). Given the wide differences in forecast windows (24/6 hours and 10-30 minutes) and SEP event energies (10-100 MeV and 300 MeV), the observational data suitable for supporting the corresponding requirement can be very different.

---

[5] https://srag.jsc.nasa.gov/SRAGFiles/Publications/SRAG_Model_Requirements.pdf
[6] https://www.swpc.noaa.gov/noaa-scales-explanation

Another important point raised was related to evaluating the improvements/advances in SEP forecasting, which is a fundamental aspect of understanding the progress in the domain and enhancing operation performance. While metrics such as True Skill Statistics (TSS) and Heidke Skill Score (HSS) have been used for forecast evaluation, these alone do not sufficiently describe the relative performance of SEP forecasting models with respect to each other. The absence of a proper evaluation of models using the same benchmark data set can result in an incompatibility to compare performances. Nevertheless, there are attempts to unify certain forecasting conditions among the efforts, such as the SEP scoreboard maintained by the CCMC, the SEP validation challenge, and related SEP Model Validation Infrastructure SPHINX conducted by NASA SRAG.

Several challenges in the existing capabilities for predicting the occurrence, properties, and evolution of SEP events were identified. These include the disconnect between current SEP research prediction efforts and operational forecasting due to differences in requirements and data streams, the critical role of data latency in SEP forecasting, and the necessity of robust real-time data to ensure accurate predictions. Additionally, the lack of real-time data for certain key parameters, such as solar radio spectrograms, further complicates operational forecasting. A coordinated community effort is required to address these challenges and enable an efficient transition of existing and emerging capabilities for the operational forecast of SEP events and associated impacts.

## IV. Current Observational Capabilities and Data Preparation Efforts.

For the development of new and improvement of the existing SEP forecast approaches, it is essential to have a complete picture of available observational capabilities and modeling, as well as utilized data analysis techniques and specialized catalogs, such as processing of daily proton flux time series (Ali et al., 2024), solar photospheric magnetic field features of ARs (Kasapis et al., 2022, Kosovich et al. 2024), and SEP event properties (e.g., Robinson et al., 2018, Rotti et al., 2022,). In particular, the discussion on space weather and SEP environment monitoring with the Solar and Heliospheric Observatory (SOHO), Solar Terrestrial Relations Observatory (STEREO), Geostationary Operational Environmental Satellite (GOES), Solar Orbiter, Radiation Assessment Detector (RAD) onboard a Martian rover Curiosity, etc., along with new mission proposals such as Earth-Moon L4 missions allow us to oversee an observational inventory for the current and future capabilities to predict SEP events and variation of the radiation environment, and track their evolution. This discussion was extended to ongoing efforts to monitor the near-Earth radiation environment, including measurements of cosmic rays and geomagnetic activity, as well as direct measurements and data mining efforts for the Automated Radiation Measurements for Aerospace Safety (ARMAS) experiment and space radiation data integration and visualization by RadLab. The general overview of the observations was complemented by the analysis of individual SEP events (e.g., extreme activity in May 2024) data processing efforts, including data benchmark preparation and comparison of SEP data sets, development of catalogs of solar filaments, and properties of the global solar activity.

A question raised during the data-related discussion (paraphrased) was: "What is the one data stream that would most improve SEP forecasting?" Several ideas were suggested, including a proposal for a mission that could detect SEPs 'upstream' (i.e. somewhere between the Sun and the location of interest along the connecting magnetic field line) and provide earlier warnings. The most intense discussion was conducted around the potential of L4 and L5 space missions, such as an upcoming Vigil mission to L5. Specifically, it was noted that the shortest pathways for travel to Mars are located around L4, and the corresponding monitoring missions at that location could play a role in safeguarding astronauts on Mars, as well as on the Moon, since SEP events are often spatially wide. The discussion also explored the possibility of predicting SEP events at the L5 point using data from Earth (as this experiment could allow us to understand the benefits of an L4 mission for SEP prediction for Earth's environment). However, this was not thought to be directly possible due to the limited number of SEP events observed near L5 as measured by STEREO satellites.

Since the majority of SEPs are assumed to be accelerated at the CME shock fronts, CME data represents an important stream for SEP forecasting (Rotti et al., 2023). The quality of CME observational data and, therefore, the constraints to the geometrical properties of CMEs and their shocks are critical. The need for more detailed and accurate CME data closer to the Sun was stressed, as these data are crucial for understanding SEP acceleration and propagation often happening close to the Sun. In particular, the near-Sun CME observations and related CME alerts by the ground-based K-Cor coronagraph could be very beneficial for improving SEP forecasting.

Another key discussion topic pertained to the solar photospheric magnetic field measurements. On the one hand, magnetic fields in ARs are only indirectly related to the SEP acceleration process as they eventually provide free energy to the particles. On the other hand, many models rely on photospheric magnetic field data (Whitman et al., 2023). In the case of the fast-arriving SEP event and the presence of data latency with respect to the coronagraph and flare data, magnetic fields remain one of the few data sources to utilize for forecasting. 24-hour All-Clear models can also rely on magnetic field information since the drivers (flares and CMEs) are not yet visible.

Magnetic field observations continue to be highly important for SEP forecasting and space weather in general. In this context, the advantages of having photospheric magnetic field measurements at an off-Sun-Earth line – such as such as those provided by Solar Orbiter or expected from the scheduled Vigil mission at L5 – was discussed. Utilizing far-side helioseismology is also important to understanding the evolution of the far-side magnetic field and detecting newly emerging active regions and, therefore, the long-term SEP forecasting.

This critical need for observations from different vantage points cannot be discussed without moving to spacecraft technology development, for instance, micropropulsion systems for long-term operational efficiency. The discussion included the transition from single-point to multi-point (constellation) observations that could lead to a significant improvement in data coverage and allow studies related to small-scale inhomogeneities in SEPs. Additionally, onboard high-level intelligent processing, largely driven by the latest developments in data science, could support real-time data processing and reduce the data volumes to be transmitted to the ground.

### V. Modeling and Forecasting Capabilities Highlights.

Modeling the propagation of the SEPs in the inner heliosphere and forecasting the expectation of an SEP event and its properties at a given spatial location remain among the key priorities for space weather science. A recent review by Whitman et al. (2023) summarizes various approaches for the SEP modeling in place, including empirical (e.g., Richardson et al., 2018, Falconer et al., 2011), physics-based (e.g., Li et al., 2021), and ML-based (e.g., Aminalragia-Giamini et al., 2021, Torres et al., 2022). We have discussed the latest developments in these approaches, including advances in physics-based models that simulate particle acceleration at CME shocks and heliospheric transport, new empirical/statistical methods for SEP prediction, and comprehensive ML-driven models, often featuring custom ML methodologies refined for this problem.

Considering the proliferation of machine learning (ML) algorithms, one of the critical questions is: "Can we trust machine learning for SEP event forecasting?" Evaluating the reliability of ML-driven results is crucial for drawing meaningful conclusions and ensuring the interpretability of ML model outputs. Currently, there is an ongoing community effort to make the results of ML and DL models interpretable. The interpretability of certain classes of ML models, however, is viewed in different ways among different stakeholders in the SEP forecasting pipeline. For instance, a modeler—who understands the internal structure of the ML model—may interpret it differently from a flight surgeon, who relies on the model's outputs to make critical decisions. The simplicity of the ML model aids its explainability, and it could be useful to have a simpler model in favor of understanding the decisions. Ensuring strong expertise across space physics and computer science teams is essential, as is the use of benchmark datasets—making SEP data 'ML-ready'—to enable proper model validation. It was also noted that benchmark datasets, along with model performance on these datasets, could serve as a foundation for building trust in machine learning models.

The common challenges of evaluating SEP prediction models, particularly given the rare nature of SEPs (strongly class-imbalanced problem), make formulating standard metrics to assess a model solution challenging. This limits the use of traditional metrics like accuracy and requires new metrics to estimate the impact of false positives and false negatives, in addition to widely used TSS and HSS. The audience mentioned the importance of considering the specific needs and priorities of end-users when selecting evaluation metrics. Along the class-imbalance line, it is important to avoid biased sampling of the data for training ML models.

All-Clear predictions are important for mission and launch planning, as well as EVAs. This question warranted a separate discussion during the workshop. The All-Clear problem is posed as confidently predicting the absence of SEP events for a certain period, allowing for safe operations in lack of atmosphere. However, the need for extremely high accuracy in such predictions was expressed, as any missed event could have acutely adverse consequences for the operators. It was also discussed whether All-Clear is just an inverse problem to the prediction of SEPs or as is a principally new setup of a problem. A possible answer may depend on the level of information one needs to make a decision. A concept comprising flare/CME predictions for an All-Clear assessment has been proposed by Georgoulis (2024).

The scarcity of SEP events remains a significant challenge, according to participants, raising questions about whether ML is an appropriate approach given its reliance on large data volumes and statistical significance, particularly for high energy and flux thresholds. It was emphasized that ML could be used beyond direct forecasting for use in synthetic data generation. Synthetic data can augment the existing data sets of SEPs and, therefore, provide more sampling for training. A previous effort by NCAR, which successfully generated SEP events (Baydin et al., 2023), was cited as an example of how synthetic data generation could support ML model training by creating diverse scenarios. An alternative to ML/DL models is physics-based ones; however, these models also can have unknown elements such as cross-field diffusion coefficients or seed particle populations for the SEP-related cases. The ML can also work synergistically with physics-based simulations, such as in ML-driven surrogate models or when using large simulation grids to learn the statistical dependences of SEP propagation.

## VI. Conclusion

Mitigating the risks posed by SEP events require a comprehensive approach that integrates monitoring, modeling, and forecasting capabilities and involves researchers, data providers, and decision-makers. The SEP workshop at Georgia State University provided valuable insights into an operational treatment of SEPs and research advancements necessary for improving space weather predictions. Key discussions focused on identifying user needs and requirements, enhancing data availability and defining data needs, and understanding better the capabilities of data-driven and physics-based forecasting techniques to improve forecasting accuracy. We believe that the workshop and related discussion supported the global space exploration goals and contributed to the advancement of our understanding of the Sun-Earth interconnected system and related space weather conditions. Additionally, by engaging the representatives from exploration, the workshop facilitated more discussions between NASA's research (SMD) and space technology (STMD) mission directorates, ensuring that research efforts are effectively coordinated with technological and operational needs. Continued collaboration between research and operational communities will be essential for advancing our ability to predict SEP events and mitigate their impact.

## Acknowledgments

We would like to thank all participants and presenters of the "Operational and Exploration Requirements and Research Capabilities for SEP Environment Monitoring and Forecasting" workshop whose active participation and expertise significantly contributed to the discussion topics. The workshop was supported by NASA: NASA ROSES 2023 program "F.2 Topical Workshops, Symposia, and Conferences", grant number 80NSSC24K1510. We are also thankful to the Metcalf Travel Award committee for providing support for a Metcalf lecturer for the workshop.